\documentclass[a4paper]{article}

\usepackage{INTERSPEECH2020}

\title{ECAPA-TDNN: Emphasized Channel Attention, Propagation and Aggregation in TDNN Based Speaker Verification}
\name{Brecht Desplanques, Jenthe Thienpondt, Kris Demuynck}

\address{
    IDLab, Department of Electronics and Information Systems, imec - Ghent University, Belgium
}
\email{brecht.desplanques@ugent.be, jenthe.thienpondt@ugent.be}

\begin{document}

\maketitle
\begin{abstract}
Current speaker verification techniques rely on a neural network to extract speaker representations. The successful x-vector architecture is a Time Delay Neural Network (TDNN) that applies statistics pooling to project variable-length utterances into fixed-length speaker characterizing embeddings. In this paper, we propose multiple enhancements to this architecture based on recent trends in the related fields of face verification and computer vision. Firstly, the initial frame layers can be restructured into 1-dimensional Res2Net modules with impactful skip connections. Similarly to SE-ResNet, we introduce Squeeze-and-Excitation blocks in these modules to explicitly model channel interdependencies. The SE block expands the temporal context of the frame layer by rescaling the channels according to global properties of the recording. Secondly, neural networks are known to learn hierarchical features, with each layer operating on a different level of complexity. To leverage this complementary information, we aggregate and propagate features of different hierarchical levels. Finally, we improve the statistics pooling module with channel-dependent frame attention. This enables the network to focus on different subsets of frames during each of the channel's statistics estimation. The proposed ECAPA-TDNN architecture significantly outperforms state-of-the-art TDNN based systems on the VoxCeleb test sets and the 2019 VoxCeleb Speaker Recognition Challenge.

\end{abstract}

\section{Introduction}

In recent years, x-vectors~\cite{x_vectors} and their subsequent improvements~\cite{x_vector_wide,jhu_voxsrc, but_voxsrc} have consistently provided state-of-the-art results on the task of speaker verification. Improving upon the original Time Delay Neural Network (TDNN) architecture is an active area of research.
Usually, the neural networks are trained on the speaker identification task. After convergence, low dimensional speaker embeddings can be extracted from the bottleneck layer preceding the output layer to characterize the speaker in the input recording.
Speaker verification can be accomplished by comparing the two embeddings corresponding with an enrollment and a test recording to accept or reject the hypothesis that both recordings contain the same speaker. A simple cosine distance measurement can be used for this comparison. In addition, a more complicated scoring backend can be trained such as Probabilistic Linear Discriminant Analysis (PLDA)~\cite{plda}.

The rising popularity of the x-vector system has resulted in significant architectural improvements and optimized training procedures~\cite{arcface} over the original approach.
The topology of the system was improved by incorporating elements of the popular ResNet~\cite{resnet} architecture. Adding residual connections between the frame-level layers has been shown to enhance the embeddings~\cite{jhu_voxsrc, but_voxsrc}. Additionally, residual connections enable the back-propagation algorithm to converge faster and help avoid the vanishing gradient problem~\cite{resnet}. 

The statistics pooling layer in the x-vector system projects the variable-length input into a fixed-length representation by gathering simple statistics of hidden node activations across time. The authors in~\cite{att_stat, self_att} introduce a temporal self-attention system to this pooling layer which allows the network to only focus on frames it deems important. It can also be interpreted as a Voice Activity Detection (VAD) preprocessing step to detect the irrelevant non-speech frames.

In this work we propose further architectural enhancements to the TDNN architecture and statistics pooling layer. We introduce additional skip connections to propagate and aggregate channels throughout the system. Channel attention that uses a global context is incorporated in the frame layers and statistics pooling layer to improve the results even further.

The paper is organized as follows: Section 2 will describe the current state-of-the-art speaker recognition systems which will be used as a baseline. Section 3 will explain and motivate the novel components of our proposed architecture. Section 4 will explain our experimental setup to test the impact of the individual components in our architecture on the popular VoxCeleb datasets~\cite{vox1,vox2,voxsrc}. We discuss the results of these experiments in Section 5. In addition, a comparison between popular state-of-the-art baseline systems will be provided. Section 6 will conclude with a brief overview of our findings.



\section{DNN speaker recognition systems}
\label{sec:baselines}

Two types of DNN-based speaker recognition architectures will serve as strong baselines to measure the impact of our proposed architecture: an x-vector and a ResNet based system which both currently provide state-of-the-art performance on speaker verification tasks such as VoxSRC~\cite{voxsrc}.

\subsection{Extended-TDNN x-vector}
\label{sec:ETDNN_baseline}

The first baseline system is the Extended TDNN x-vector architecture~\cite{x_vector_wide,jhu_voxsrc, but_voxsrc} and improves upon the original x-vector system introduced in~\cite{x_vectors}. The initial frame layers consist of 1-dimensional dilated convolutional layers interleaved with dense layers. Every filter has access to all the features of the previous layer or input layer. The task of the dilated convolutional layers is to gradually build up the temporal context. Residual connections are introduced in all frame-level layers. The frame layers are followed by an attentive statistics pooling layer that calculates the mean and standard deviations of the final frame-level features. The attention system~\cite{att_stat} allows the model to select the frames it deems relevant. After the statistics pooling, two fully-connected layers are introduced with the first one acting as a bottleneck layer to generate the low-dimensional speaker characterizing embedding.

\subsection{ResNet-based r-vector}
\label{sec:resnet_baseline}

The second baseline system is the r-vector system proposed in~\cite{but_voxsrc}.
It is based on the ResNet18 and ResNet34 implementations of the successful ResNet architecture~\cite{resnet}. The convolutional frame layers of this network process the features as a 2-dimensional signal before collecting the mean and standard deviation statistics in the pooling layer. See~\cite{but_voxsrc} for more details about the topology.

\section{Proposed ECAPA-TDNN architecture}
\label{sec:prop_architecture}

In this section, we examine some of the limitations of the x-vector architecture and incorporate potential solutions in our ECAPA-TDNN architecture. The following subsections will focus on frame-level and pooling-level enhancements. An overview of the complete architecture is given by Figure \ref{fig:full_network}. BN stands for Batch Normalization~\cite{batch_norm} and the non-linearities are Rectified Linear Units (ReLU) unless specified otherwise.

\subsection{Channel- and context-dependent statistics pooling}
\label{cas}

In recent x-vector architectures, soft self-attention is used for calculating weighted statistics in the temporal pooling layer~\cite{att_stat}. Success with multi-headed attention has shown that certain speaker properties can be extracted on different sets of frames~\cite{self_att}. Due to these results, we argue that it might be beneficial to extend this temporal attention mechanism even further to the channel dimension. This enables the network to focus more on speaker characteristics that do not activate on identical or similar time instances, e.g.\ speaker-specific properties of vowels versus speaker-specific properties of consonants. 

We implement the attention mechanism as described in \cite{att_stat} and adapt it to be channel-dependent:
\begin{equation}
\label{eq1_stat_pool}
e_{t,c} = \pmb{v}^{T}_{c} f(\pmb{W}\pmb{h}_{t} + \pmb{b}) + k_{c},
\end{equation}
where $\pmb{h}_{t}$ are the activations of the last frame layer at time step $t$. The parameters $\pmb{W} \in \mathbb{R}^{R \times C}$ and $\pmb{b} \in \mathbb{R}^{R \times 1}$ project the information for self-attention into a smaller $R$-dimensional representation that is shared across all $C$ channels to reduce the parameter count and risk of overfitting. After a non-linearity $f(\cdot)$ this information is transformed to a channel-dependent self-attention score through a linear layer with weights $\pmb{v}_{c} \in \mathbb{R}^{R \times 1}$ and bias $k_{c}$. This scalar score $e_{t,c}$ is then normalized over all frames by applying the softmax function channel-wise across time:
\begin{equation}
\label{eq2_stat_pool}
\alpha_{t,c} = \frac{\exp(e_{t,c})}{\sum_{\tau}^{T} \exp(e_{\tau,c})}.
\end{equation}

The self-attention score $\alpha_{t,c}$ represents the importance of each frame given the channel and is used to calculate the weighted statistics of channel $c$. For each utterance the channel component $\tilde{\mu}_{c}$ of the weighted mean vector $\pmb{\tilde{\mu}}$ is estimated as:
\begin{equation}
\label{eq3_stat_pool}
\tilde{\mu}_{c} = \sum_{t}^{T}\alpha_{t,c}h_{t,c}.
\end{equation}
The channel component $\tilde{\sigma}_{c}$ of the weighted standard deviation vector $\pmb{\tilde{\sigma}}$ is constructed as follows:
\begin{equation}
\label{eq4_stat_pool}
\tilde{\sigma}_{c} = \sqrt{\sum_{t}^{T} \alpha_{t,c}h_{t,c}^{2}-\tilde{\mu}_{c}^{2}}.
\end{equation}
The final output of the pooling layer is given by concatenating the vectors of the weighted mean $\pmb{\tilde{\mu}}$ and weighted standard deviation $\pmb{\tilde{\sigma}}$.

Furthermore, we expand the temporal context of the pooling layer by allowing the self-attention to look at global properties of the utterance. We concatenate the local input $\pmb{h}_{t}$ in \eqref{eq1_stat_pool} with the global non-weighted mean and standard deviation of $\pmb{h}_{t}$ across the time domain. This context vector should allow the attention mechanism to adapt itself to global properties of the utterance such as noise or recording conditions.

\subsection{1-Dimensional Squeeze-Excitation Res2Blocks} \label{sec:se_block}

The temporal context of frame layers in the original x-vector system is limited to 15 frames. As the network apparently benefits from a wider temporal context \cite{x_vector_wide, but_voxsrc, jhu_voxsrc} we argue it could be beneficial to rescale the frame-level features given global properties of the recording, similar to the global context in the attention module described above. For this purpose we introduce 1-dimensional Squeeze-Excitation (SE) blocks, as this computer vision approach to model global channel interdependencies has been proved successful \cite{se_block, se_speech}.

The first component of an SE-block is the squeeze operation which generates a descriptor for each channel. The squeeze operation simply consists of calculating the mean vector $\pmb{z}$ of the frame-level features across the time domain:
\begin{equation}
\label{se_squeeze}
\pmb{z} = \frac{1}{T}\sum_{t}^{T}\pmb{h}_t.
\end{equation}
The descriptors in $\pmb{z}$ are then used in the excitation operation to calculate a weight for each channel. We define the subsequent excitation operation as:
\begin{equation}
\label{se_excitate}
\pmb{s} = \sigma( \pmb{W}_2 f(\pmb{W}_1 \pmb{z} + \pmb{b}_{1}) + \pmb{b}_2)
\end{equation}
with $\sigma(\cdot)$ denoting the sigmoid function, $f(\cdot)$ a non-linearity, $\pmb{W}_{1} \in \mathbb{R}^{R \times C}$ and $\pmb{W}_{2} \in \mathbb{R}^{C \times R}$. This operation acts as a bottleneck layer with $C$ and $R$ referring to the number of input channels and reduced dimensionality respectively. The resulting vector $\pmb{s}$ contains weights $s_c$ between zero and one, which are applied to the original input through channel-wise multiplication
\begin{equation}
    \tilde{\pmb{h}}_{c} = s_{c}\pmb{h}_{c}
    \label{se_final}
\end{equation}

The 1-dimensional SE-block can be integrated in the x-vector architecture in various ways, with using them after each dilated convolution being the most straightforward one. However, we want to combine them with the benefits of residual connections~\cite{resnet}. Simultaneously, we do not want to increase the total amount of parameters too much compared to the baseline systems.
\begin{figure}[ht]
\centering
\includegraphics[scale=0.26]{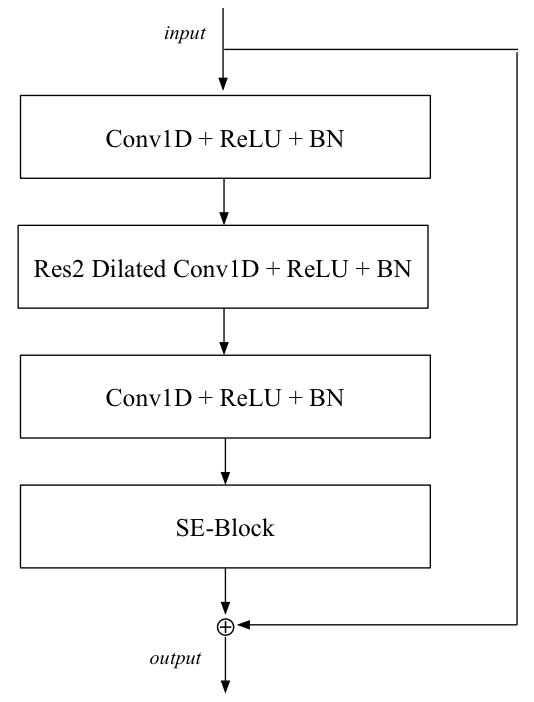}
\caption{{\it The SE-Res2Block of the ECAPA-TDNN architecture. The standard Conv1D layers have a kernel size of 1. The central Res2Net~\cite{res2net} Conv1D with scale dimension $s=8$ expands the temporal context through kernel size $k$ and dilation spacing $d$.}}
\label{fig:se_resblock}
\end{figure}
The \textit{SE-Res2Block} shown in Figure \ref{fig:se_resblock} incorporates the requirements mentioned above.
We contain the dilated convolutions with a preceding and succeeding dense layer with a context of 1 frame. The first dense layer can be used to reduce the feature dimension, while the second dense layer restores the number of features to the original dimension. This is followed by an SE-block to scale each channel. The whole unit is covered by a skip connection.

The use of these traditional ResBlocks makes it easy to incorporate advancements concerning this popular computer vision architecture. The recent Res2Net module~\cite{res2net}, for example, enhances the central convolutional layer so that it can process multi-scale features by constructing hierarchical residual-like connections within. The integration of this module improves performance, while significantly reducing the number of model parameters.

\subsection{Multi-layer feature aggregation and summation}

The original x-vector system only uses the feature map of the last frame-layer for calculating the pooled statistics. Given the hierarchical nature of a TDNN, these deeper level features are the most complex ones and should be strongly correlated with the speaker identities. However, due to evidence in~\cite{mfa_tag,multi_stage_aggregation} we argue that the more shallow feature maps can also contribute towards more robust speaker embeddings. For each frame, our proposed system concatenates the output feature maps of all the SE-Res2Blocks. After this Multi-layer Feature Aggregation (MFA), a dense layer processes the concatenated information to generate the features for the attentive statistics pooling.

\begin{figure}[t]
\centering
\includegraphics[scale=0.28]{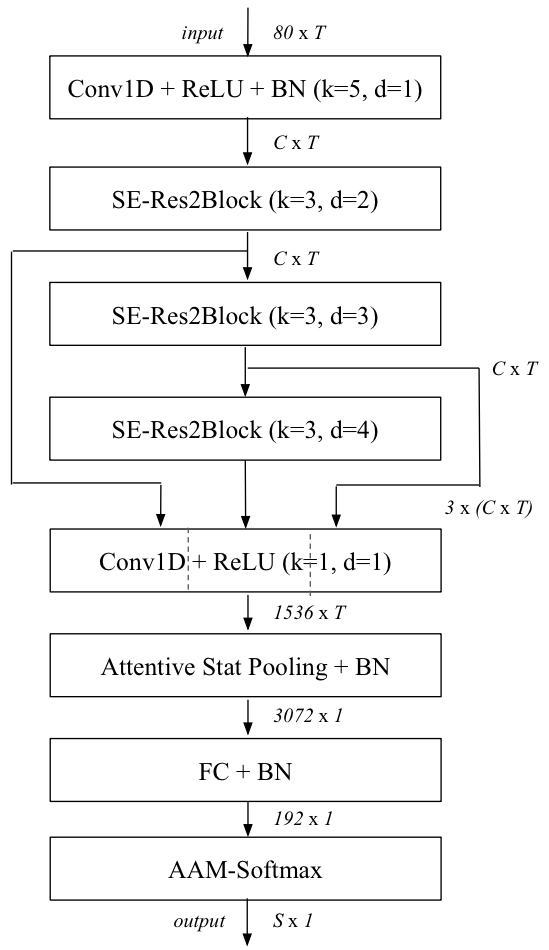}
\caption{{Network topology of the ECAPA-TDNN. We denote $k$ for kernel size and $d$ for dilation spacing of the Conv1D layers or SE-Res2Blocks. $C$ and $T$ correspond to the channel and temporal dimension of the intermediate feature-maps respectively. $S$ is the number of training speakers.}}
\label{fig:full_network}
\end{figure}

Another, complementary way to exploit multi-layer information is to use the output of all preceding SE-Res2Blocks and initial convolutional layer as input for each frame layer block~\cite{mfa_tag,multi_stage_summation}. We implement this by defining the residual connection in each SE-Res2Block as the sum of the outputs of all the previous blocks. We opt for a summation of the feature maps instead of concatenation to restrain the model parameter count. The final architecture without the summed residual connections is shown in Figure~\ref{fig:full_network}.

\begin{table*}
  \caption{EER and MinDCF performance of all systems on the standard VoxCeleb1 and VoxSRC 2019 test sets.}
  \label{tab:exp_results}
  \centering
  \begin{tabular}{ccccccccc}
    \toprule
    \multicolumn{1}{c}{\textbf{Architecture}} &
    \multicolumn{1}{c}{\textbf{\# Params}} &
    \multicolumn{2}{c}{\textbf{VoxCeleb1}} &
    \multicolumn{2}{c}{\textbf{VoxCeleb1-E}} & 
    \multicolumn{2}{c}{\textbf{VoxCeleb1-H}} &
    \multicolumn{1}{c}{\textbf{VoxSRC19}} \\
    \cmidrule(lr){3-4} \cmidrule(lr){5-6} \cmidrule(lr){7-8} \cmidrule(lr){9-9}
    \multicolumn{2}{c}{\textbf{}} & 
    \multicolumn{1}{c}{\textbf{EER(\%)}} & \multicolumn{1}{c}{\textbf{MinDCF}} &
    \multicolumn{1}{c}{\textbf{EER(\%)}} & \multicolumn{1}{c}{\textbf{MinDCF}} &
    \multicolumn{1}{c}{\textbf{EER(\%)}} & \multicolumn{1}{c}{\textbf{MinDCF}} &
    \multicolumn{1}{c}{\textbf{EER(\%)}} \\
    \midrule
    E-TDNN & 6.8M & 1.49 & 0.1604 & 1.61  & 0.1712 & 2.69 & 0.2419 & 1.81\\
    E-TDNN (large) & 20.4M & 1.26 & 0.1399 & 1.37 & 0.1487 & 2.35 & 0.2153 & 1.61 \\
    \midrule
    ResNet18 & 13.8M & 1.47 & 0.1772 & 1.60 & 0.1789 & 2.88 & 0.2672 & 1.97 \\
    ResNet34 & 23.9M & 1.19 & 0.1592 & 1.33 & 0.1560 & 2.46 & 0.2288 & 1.57  \\
    \midrule
    \midrule
    ECAPA-TDNN (C=512) & 6.2M & 1.01 & 0.1274 & 1.24 & 0.1418 & 2.32 & 0.2181 & 1.32 \\
    ECAPA-TDNN (C=1024) & 14.7M & \textbf{0.87} & \textbf{0.1066} & \textbf{1.12} & \textbf{0.1318} & \textbf{2.12} & \textbf{0.2101} & \textbf{1.22} \\
    \bottomrule
  \end{tabular}
\end{table*}

\section{Experimental setup}


\subsection{Training the speaker embedding extractors}


We apply the fixed-condition VoxSRC 2019 training restrictions \cite{voxsrc} and only use the development part of the VoxCeleb2 dataset~\cite{vox2} with 5994 speakers as training data. A small subset of about 2\% of the data is reserved as a validation set for hyperparameter optimization. It is a well known fact that neural networks benefit from data augmentation which generates extra training samples. We generate a total of 6 extra samples for each utterance. The first set of augmentations follow the Kaldi recipe~\cite{x_vector_wide} in combination with the publicly available MUSAN dataset (babble, noise) \cite{musan} and the RIR dataset (reverb) provided in~\cite{rirs}. The remaining three augmentations are generated with the open-source SoX (tempo up, tempo down) and FFmpeg (alternating opus or aac compression) libraries.

The input features are 80-dimensional MFCCs from a 25~ms window with a 10 ms frame shift. Two second random crops of the MFCCs feature vectors are normalized through cepstral mean subtraction and no voice activity detection is applied. As a final augmentation step, we apply SpecAugment~\cite{specaugment} on the log mel spectrogram of the samples. The algorithm randomly masks 0 to 5 frames in the time domain and 0 to 10 channels in the frequency domain.

All models are trained with a cyclical learning rate varying between 1e-8 and 1e-3 using the \textit{triangular2} policy as described in \cite{clr} in conjunction with the Adam optimizer \cite{adam}. The duration of one cycle is set to 130k iterations. All systems are trained using AAM-softmax \cite{arcface,aam_embeddings} with a margin of 0.2 and softmax prescaling of 30 for 4 cycles. To prevent overfitting, we apply a weight decay on all weights in the model of 2e-5, except for the AAM-softmax weights, which uses 2e-4. The mini-batch size for training is 128.

We study two setups of the proposed ECAPA-TDNN architecture with either 512 or 1024 channels in the convolutional frame layers. The dimension of the bottleneck in the SE-Block and the attention module is set to 128. The scale dimension $s$ in the Res2Block~\cite{res2net} is set to 8. The number of nodes in the final fully-connected layer is 192. The performance of this system will be compared to the baselines described in Section~\ref{sec:baselines}.

\subsection{Speaker verification}

Speaker embeddings are extracted from the final fully-connected layer for all systems. Trial scores are produced using the cosine distance between embeddings. Subsequently, all the scores are normalized using adaptive s-norm~\cite{as_norm_original_1, as_norm_original_2}. The imposter cohort consists of the speaker-wise averages of the length-normalized embeddings of all training utterances. The size of the imposter cohort was set to 1000 for the VoxCeleb test sets and to a more robust value of 50 for the cross-dataset VoxSRC 2019 evaluation.

\subsection{Evaluation protocol}

The system is evaluated on the popular VoxCeleb1 test sets~\cite{vox1} and VoxSRC 2019 evaluation set~\cite{voxsrc}. Performance will be measured by providing the Equal Error Rate (EER) and the minimum normalized detection cost MinDCF with $P_{target}=10^{-2}$ and $C_{FA}=C_{Miss}=1$. A concise ablation study is used to gain a deeper understanding how each of the proposed improvements affects the performance.

\section{Results}


A performance overview of the baseline systems described in Section~\ref{sec:baselines} and our proposed ECAPA-TDNN system is given in Table \ref{tab:exp_results}, together with the number of model parameters in the embedding extractor. We implement two setups with the number of filters $C$ in the convolutional layers either set to 512 or 1024. Our proposed architecture significantly outperforms all baselines while using fewer model parameters. The larger ECAPA-TDNN system gives an average relative improvement of 18.7\% in EER and 12.5\% in MinDCF over the best scoring baseline for each test set. We note that the performance of the baselines supersedes the numbers reported in~\cite{jhu_voxsrc, but_voxsrc} in most cases. We continue with an ablation study of the individual components introduced in Section~\ref{sec:prop_architecture}. An overview of these results is given in Table \ref{tab:ablation}.

\begin{table}[h]
  \caption{Ablation study of the ECAPA-TDNN architecture.}
  \label{tab:ablation}
  \centering
  \begin{tabular}{cccc}
    \toprule
     & \textbf{Systems} & \multicolumn{1}{c}{\textbf{EER(\%)}} & \multicolumn{1}{c}{\textbf{MinDCF}} \\
    \midrule
     & \hfill ECAPA-TDNN (C=512) & 1.01 & 0.1274 \\
    \midrule
    \midrule
    A.1 &  \hfill Attentive Statistics~\cite{att_stat} & 1.12 & 0.1316 \\
    A.2 & \hfill Channel Att. w/o Context & 1.03 & 0.1288 \\
    \midrule
    B.1 & \hfill No SE-Block & 1.27 & 0.1446 \\
    B.2 & \hfill No Res2Net-Block & 1.07 & 0.1316 \\
    \midrule
    C.1 & \hfill No MFA & 1.10 & 0.1311 \\
    C.2 & \hfill No Res. Connections & 1.08 & 0.1310 \\
    C.3 & \hfill No Sum Res. Connections & 1.08 & 0.1217 \\
    \bottomrule
  \end{tabular}
\end{table}


To measure the impact of our proposed attention module, we run an experiment \textit{A.1}, that uses the attention module from~\cite{att_stat}. We also run a separate experiment \textit{A.2} that does not supply the context vector to the proposed attention. The channel- and context-dependent statistics pooling system improves the EER and MinDCF metric with 9.8\% and 3.2\%, respectively. This confirms the benefits of applying different temporal attention to each channel. Addition of the context vector results in very small performance gains with the system relatively improving about 1.9\% in EER and 1.1\% in MinDCF. Nonetheless, this strengthens our belief that a TDNN-based architecture should try to exploit global context information.




This intuition is confirmed with experiment \textit{B.1} that clearly shows the importance of the SE-blocks described in Section~\ref{sec:se_block}. Incorporating the SE-modules in the Res2Blocks results in relative improvements of 20.5\% in EER and 11.9\% in the MinDCF metric. This indicates that the limited temporal context of the frame-level features is insufficient and should be complemented with global utterance-based information.
In experiment \textit{B.2} we replaced the multi-scale features of the Res2Blocks with the standard central dilated 1D convolutional of the ResNet counterpart. Aside from a substantial 30\% relative reduction in model parameters, the multi-scale Res2Net approach also leads towards a relative improvement of 5.6\% in EER and 3.2\% in MinDCF.




In experiment \textit{C.1}, we only use the output of the final SE-Res2Block instead of aggregating the information of all SE-Res2Blocks. Aggregation of the outputs leads to relative improvements of 8.2\% in EER and 2.8\% in the MinDCF value. Removing all residual connections (experiment \textit{C.2}) shows a similar rate of degradation. Replacing a standard ResNet skip connection in the SE-Res2Blocks by the sum of the outputs of all previous SE-Res2Blocks improves the EER with 6.5\%, while slightly degrading the MinDCF score in experiment \textit{C.3}. 
However, experiments during the recently held Short-duration Speaker Verification (SdSV) Challenge 2020~\cite{sdsv_plan} convinced us to incorporate summed residuals in the final ECAPA-TDNN architecture. The strong results in this challenge show the architecture generalizes well to other domains~\cite{sdsvc_paper}. 

\section{Conclusion}

In this paper we presented ECAPA-TDNN, a novel TDNN-based speaker embedding extractor for speaker verification. We built further upon the original x-vector architecture and put more Emphasis on Channel Attention, Propagation and Aggregation. The incorporation of Squeeze-Excitation blocks, multi-scale Res2Net features, extra skip connections and channel-dependent attentive statistics pooling, led to significant relative improvements of 19\% in EER on average over strong baseline systems on the VoxCeleb and VoxSRC 2019 evaluation sets.


\bibliographystyle{IEEEtran}

\bibliography{mybib}

\end{document}